\documentclass[fleqn,usenatbib]{mnras}

\usepackage{newtxtext,newtxmath}
\usepackage[T1]{fontenc}

\DeclareRobustCommand{\VAN}[3]{#2}
\let\VANthebibliography\thebibliography
\def\thebibliography{\DeclareRobustCommand{\VAN}[3]{##3}\VANthebibliography}

\usepackage{graphicx}	% Including figure files
\usepackage{amsmath}	% Advanced maths commands
\title[Decomposition of stellar populations]{Decomposition of stellar populations in CosmoDC2 galaxies using SCARLET and Deep Learning}

\author[S. Kunság-Máté et al.]{
Sándor Kunsági-Máté,$^{1}$\thanks{E-mail: skunsagimate@student.elte.hu}
István Csabai,$^{1}$
\\
$^{1}$Department of Physics of Complex Systems, ELTE Eötvös Loránd University, Pázmány Péter sétány 1/a, Budapest 1117, Hungary\\
}

\date{Accepted XXX. Received YYY; in original form ZZZ}

\pubyear{2021}

\begin{document}
\label{firstpage}
\pagerange{\pageref{firstpage}--\pageref{lastpage}}
\maketitle

% Abstract of the paper
\begin{abstract}
We are presenting a novel, Deep Learning based approach to estimate the normalized broadband spectral energy distribution (SED) of different stellar populations in synthetic galaxies. In contrast to the non-parametric multiband source separation algorithm, \texttt{SCARLET} - where the SED and morphology are simultaneously fitted - in our study we provide a \textit{morphology-independent}, statistical determination of the SEDs, where we only use the color distribution of the galaxy. We developed a neural network (\texttt{sedNN}) that accurately predicts the SEDs of the old, red and young, blue stellar populations of realistic synthetic galaxies from the color distribution of the galaxy-related pixels in simulated broadband images. We trained and tested the network on a subset of the recently published CosmoDC2 simulated galaxy catalog containing about 3,600 galaxies. The model performance was compared to the results of \texttt{SCARLET}, where we found that \texttt{sedNN} can predict the SEDs with 4-5\% accuracy on average, which is about two times better than applying \texttt{SCARLET}. We also investigated the effect of this improvement on the flux determination accuracy of the bulge and disk. We found that using more accurate SEDs decreases the error in the flux determination of the components by approximately 30\%.
\end{abstract}

% Select between one and six entries from the list of approved keywords.
% Don't make up new ones.

%%%%%%%%%%%%%%%%%%%%%%%%%%%%%%%%%%%%%%%%%%%%%%%%%%

%%%%%%%%%%%%%%%%% BODY OF PAPER %%%%%%%%%%%%%%%%%%

\section{Introduction}

\begin{figure}
	\includegraphics[width=\columnwidth]{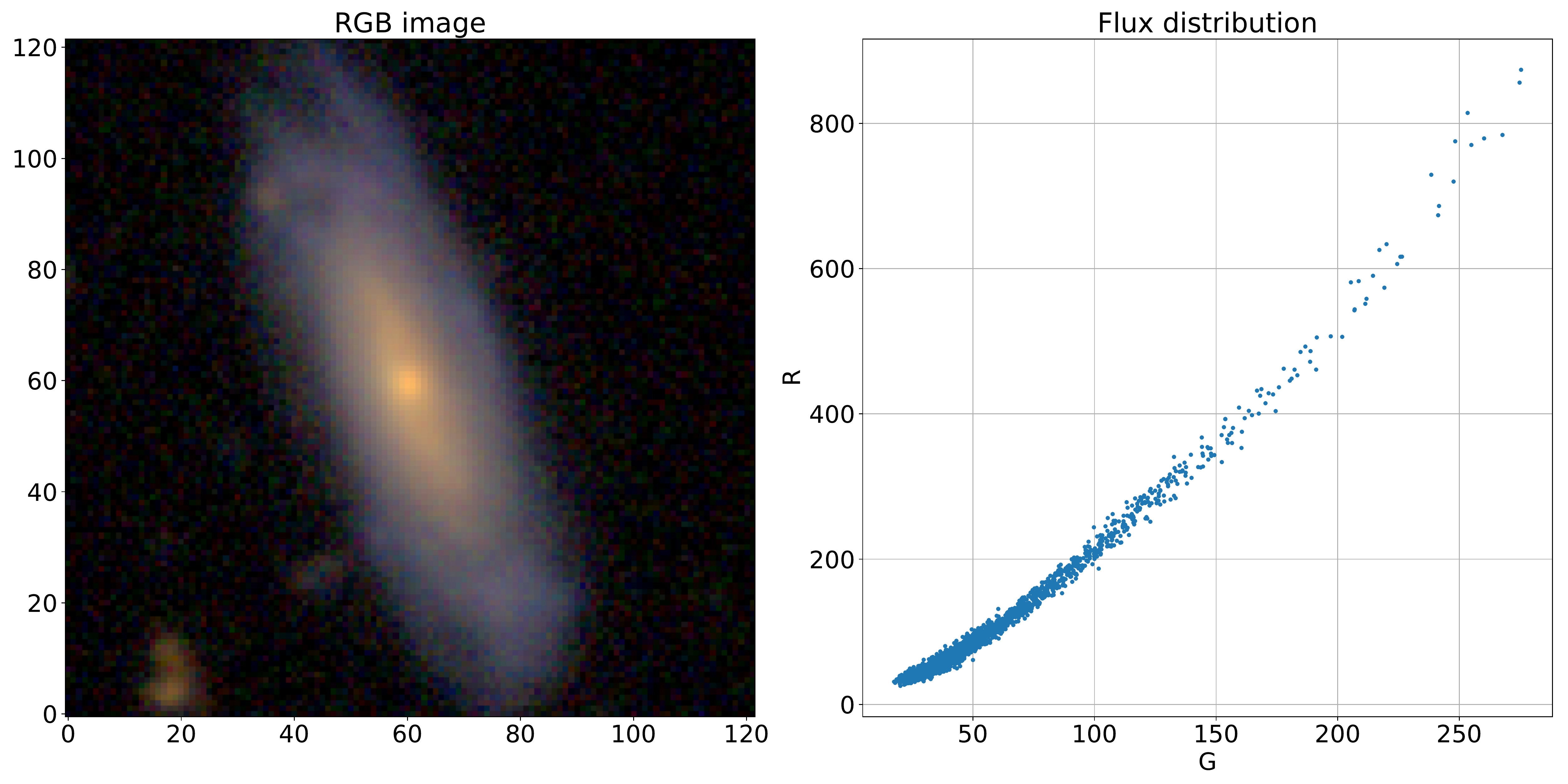}
    \caption{An example galaxy (RA: 41.422048, DEC: -9.074144) from the Dark Energy Survey (DES). \textit{Left}: RGB image of the galaxy using the $G$, $R$, $I$ filters based on the algorithm of \citet{Lupton2004}. \textit{Right}: Distribution of pixels in the $G-R$ flux space.}
    \label{fig:gal_filt_dist}
\end{figure}

The proper mechanism of the galaxy evolution is an active research area where the analysis of the bulge-disk relation and stellar population properties are crucial. Spiral galaxies have significant bimodality in their stellar composition. The disk and the spiral arms are composed of a young, blue stellar population and the central region, the bulge contains mostly old, red stars. The proper separation of the two components in multiband photometric surveys can help to understand the stellar evolution inside spiral galaxies. It is often called as a blind source separation problem, since we do not know the exact broadband spectral energy distribution (SED) and the morphology of the sources. There are two main groups of source separation algorithms: parametric and non-parametric methods.\\
In case of parametric models we assume a specific surface brightness profile for the two main components of spiral galaxies. In most cases we suppose that the disk has an exponential profile, while the bulge can be described by a de Vaucouleurs model \citet{devac}. An alternative choice is the Sersic model, in which both components can be described but with different Sersic exponents ($n=1$ for the disk, and $n=4$ for the bulge). The assumed model is then fitted to the galaxy image measured in single-band or multiband surveys. Such source separation algorithms are for example PSFex \citet{Bertin2011} or Megamorph \citet{Vika2013} fitting codes. The obvious disadvantage of this approach is that complex structures of galaxies cannot be reproduced, and we strongly rely on assumptions regarding the brightness distribution of the galaxy.\\
The other way of separating different sources from each other is to exploit the diversity of their SEDs. Let's assume that we have $N_s = 2$ sources, namely the bulge and the disk. We know that they are composed of different stellar populations, hence they are characterized by two different SEDs. This means that if we have images at $N_b \geq N_s$ distinct broadband filters, then the intensity distribution of the sources can be estimated by solving a linear system of equations. Let us denote the data observed in $N_b$ filters as $[y_i]_{i=1,...,N_b}$. It is then a linear combination of $N_s$ sources, see Equation \ref{eq:sources}.

\begin{equation}
    y_i[k] = \sum\limits_{j=1}^{N_s} a_{i,j} x_j[k] + z_i[k],
    \label{eq:sources}
\end{equation}

where $y_i[k]$ is the $k$th pixel of the original image in the $i$th band, $a_{i,j}$ and $x_j[k]$ contain the SEDs and the morphology of the sources, respectively and $z_i[k]$ is a noise term. Let's rewrite this in matrix form, see Equation \ref{eq:sources_mtx}

\begin{equation}
    Y = AX + Z,
    \label{eq:sources_mtx}
\end{equation}

where $Y$ and $Z$ are $N_b \times N_p$ matrices ($N_p$ is the number of pixels), $A$ is the SED mixing matrix and $X$ is the $N_s \times N_p$ matrix, which contains the components $x_j$. The MuSCADeT algorithm \citet{muscadet} is designed to deblend different astronomical sources having different SEDs. Its performance was demonstrated mainly in the separation of blended objects due to strong gravitational lensing, but it is also able to disentangle a single galaxy, as was shown on a spiral galaxy with significantly high colour-contrast in the original paper \citet{muscadet}. It is a multi-channel extension of the morphological component analysis (MCA) described in \citet{starck2004}. To the best of our knowledge our work is the first in applying \textit{non-parametric} source separation algorithm on a large number of synthetic galaxies to get the bulge and disk components.\\
The main challenge of this approach is that we neither know the SEDs nor the morphology of the galaxy components. There are two ways to address this problem: 1) one is to first estimate somehow the SEDs and then fix them during the morphology fitting, or 2) fit both of them and introduce some constraints to reduce degeneracy (e.g. positivity, monotonicity, symmetry constraints). MuSCADeT belongs to the first group, where the $A$ mixing matrix can be a user defined parameter, but an automated SED analysing process based on principal component analysis (PCA) is also available. This pre-processing method essentially searches for bright regions and performs PCA in the $N_b$ dimensional space spanned by the broadband filters. Hence, the pixels having proportional SEDs - meaning that they belong to the same stellar population - will be distributed along the first few PCA axes, and one can identify them using some clustering algorithm and get the mean of their SEDs. This method can be powerful, if the different stellar populations are similarly bright, both causing significant signal in the principal component analysis. The problem is that in most of the spiral galaxies there is a very bright red bulge, and some faint blue spiral arms, see Figure \ref{fig:gal_filt_dist}. In the flux distribution plot we can see that there is a very large brightness gradient, and a slight curvature of the distribution can be also observed. This curvature is due to the simultaneously changing in brightness and color from blue to red. As we can see there is no clear difference between the two stellar populations in the filter space. Therefore we had to use a different approach to determine the spectral energy distributions of the red and blue regions. Another drawback of MuSCADeT is that it does not take the PSF convolution into account during deblending.\\
A more sophisticated framework is \texttt{SCARLET} \citet{scarlet}, which is based on a generalization of non-negative matrix factorization to alternative and several simultaneous constraints. It can simultaneously fit the SEDs and the morphology, and it also applies the PSF convolution. The main focus of this method is again to disentangle different galaxies from each other, since the upcoming sky surveys (e.g. LSST \citet{lsst}) will provide images of crowded regions in the far Universe. Therefore its performance was tested on separating synthetic overlapping galaxies from each other. Hence, besides the introduction and application of our novel approach we also provide a reliable test of SCARLET on separating synthetic galaxies into the bulge and disk components to which we have compared our results.\\
Separating stellar populations inside a galaxy can be much more difficult, since the flux contribution of them is often very different (e.g. irregular galaxies with small bulge and bright star-forming regions, or older, more evolved spiral galaxies with a massive bulge and faint blue spiral arms). Moreover, in most of the cases the two stellar populations are strongly overlapping  and there is no region of the galaxy where only one of the components has a flux contribution. This fact causes high degeneracy in the SED-morphology determination and therefore there is a strong need for a morphology-independent estimation of the SEDs. In our study we have investigated simulated galaxies with different flux contributions of the bulge and disk components, and estimated their SEDs. We trained a neural network (\texttt{sedNN}) in the Keras framework that can predict the two SEDs directly from the color distribution of the galaxy.  The resulting SEDs were used as input in the morphology determination of the two sources in a TensorFlow model. This paper is organized as follows: in Section \ref{data} we introduce the synthetic data we used, in Section \ref{methods} we explain the pre-processing steps for creating the best input data as well as the architecture of our neural network, in Section \ref{results} we show our results made on the data as well as the comparison to \texttt{SCARLET} and finally in Section \ref{conclusions} we discuss and summarize the key findings of our work.

\section{Data}\label{data}

\begin{figure}
	\includegraphics[width=\columnwidth]{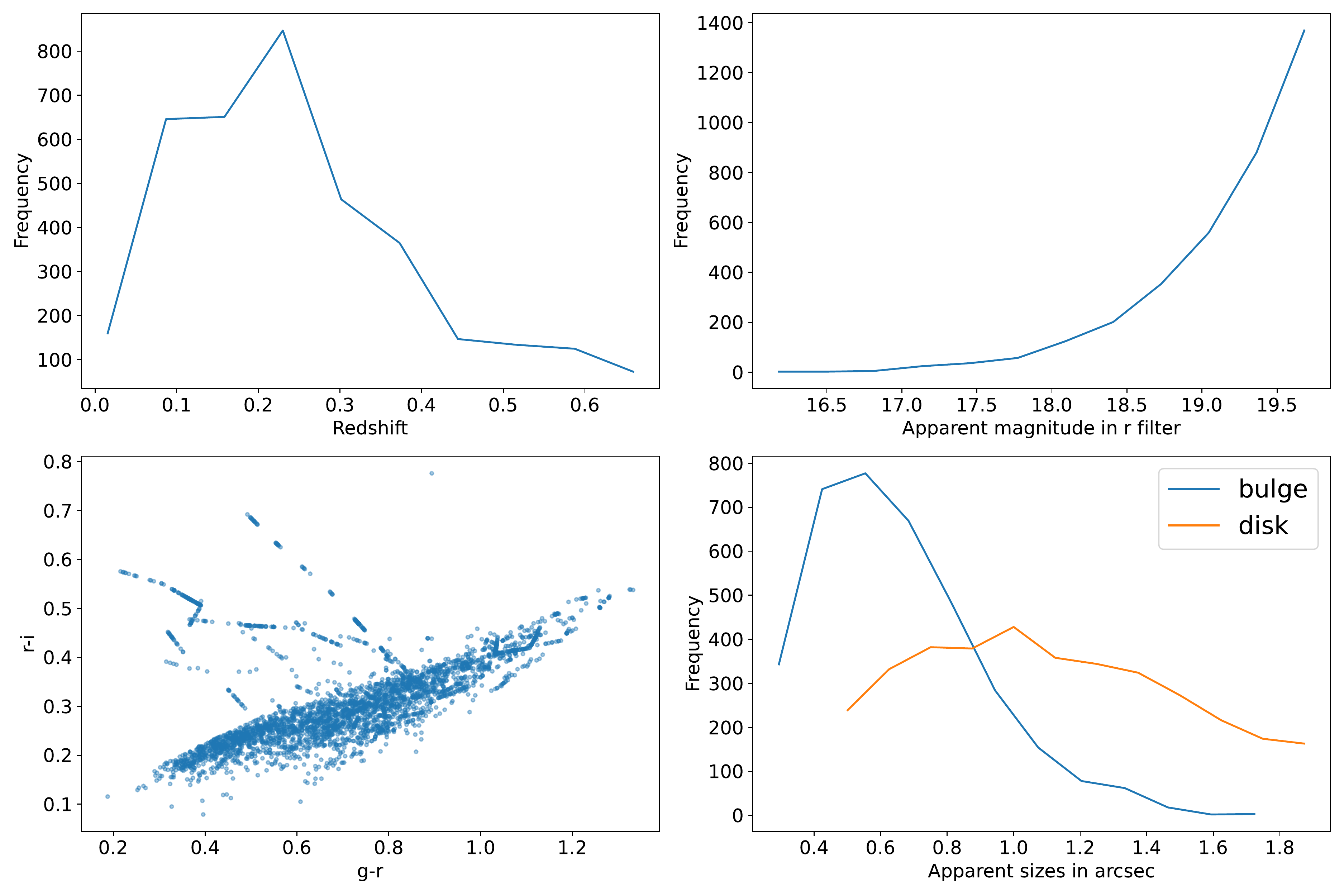}
    \caption{Distributions of important galaxy properties. \textit{Top row}: Redshift and apparent $r$ filter magnitude distribution of the galaxies. \textit{Bottom row}: $g-r$ and $r-i$ color index distribution of the galaxies as well as the apparent size distribution (half-light radius) of the bulge and disk measured in arcsec.}
    \label{fig:data_dist}
\end{figure}

We used the recently published data of CosmoDC2 synthetic sky catalog \citet{cosmodc2}, which is designed to support precision dark energy science with LSST. The catalog is based on a trillion-particle, $(4.225 Gpc)^3$ box cosmological N-body simulation. Among many properties each galaxy is characterized by a unique morphology, spectral energy distributions and broadband filter magnitudes. This catalog was constructed using empirical methods with a semi-analyitic modeling using the Galacticus code \citet{benson}. The simulated galaxies are composed of a bulge and disk both having a Sersic profile with $n=4$ and $n=1$, respectively. Galacticus follows the evolution of the components separately. The bulge is formed as a result of galaxy mergers or via instabilities of the galactic disk. Hence, we could use unbiased data of highly realistic galaxy components, where all necessary parameters were given, such as apparent magnitudes in the LSST g, r, i, z filters, shear parameters and brightness profile to describe the morphology as well as the redshifts. To get a convenient subset of galaxies we made some filtering in three parameters using 5 tiles of CosmoDC2\footnote{ref: https://portal.nersc.gov/project/lsst/cosmoDC2/} (8786, 8787, 8788, 8789 and 8790): 1) we searched for relatively bright galaxies with an apparent magnitude of LSST r band of $m_r < 20.0$; 2) we made a cut in the half light right radius ($r_{hl}$) of the disk component: $0.5 \geq r_{hl} \geq 2.0$ arcsec to avoid unresolvable galaxy components as well as hanging over from the image and 3) obviously we required exactly 2-component galaxies by setting the bulge-to-total flux ratio ($BT$) between $0.1 \geq BT \geq 0.9$. After some preliminary investigation of the filtered data we found that a huge amount of the galaxies have components with almost the same SED. Therefore we calculated the angle ($\alpha$) between the four-element SED vectors of the bulge and disk and we made a final filtering for galaxies where $\alpha > 0.1$ radian. Hence the remaining dataset consisted about 3,600 synthetic galaxies. This dataset was splitted into training, validation and test set with sizes of 2369, 539 and 704, respectively. The distributions of the most important properties of the galaxies can be seen in Figure \ref{fig:data_dist}.\\
To generate simulated LSST images of the galaxies we used the popular \texttt{GalSim} code \citet{galsim}. We have calibrated the pixel intensities ($I$) to the apparent magnitude ($m$) according to equation \ref{eq:pixint} similar to the Dark Energy Survey (DES) \citet{desdr2}.

\begin{equation}
    m = 30 - 2.5 \log_{10} I
    \label{eq:pixint}
\end{equation}

\begin{figure}
	\includegraphics[width=\columnwidth]{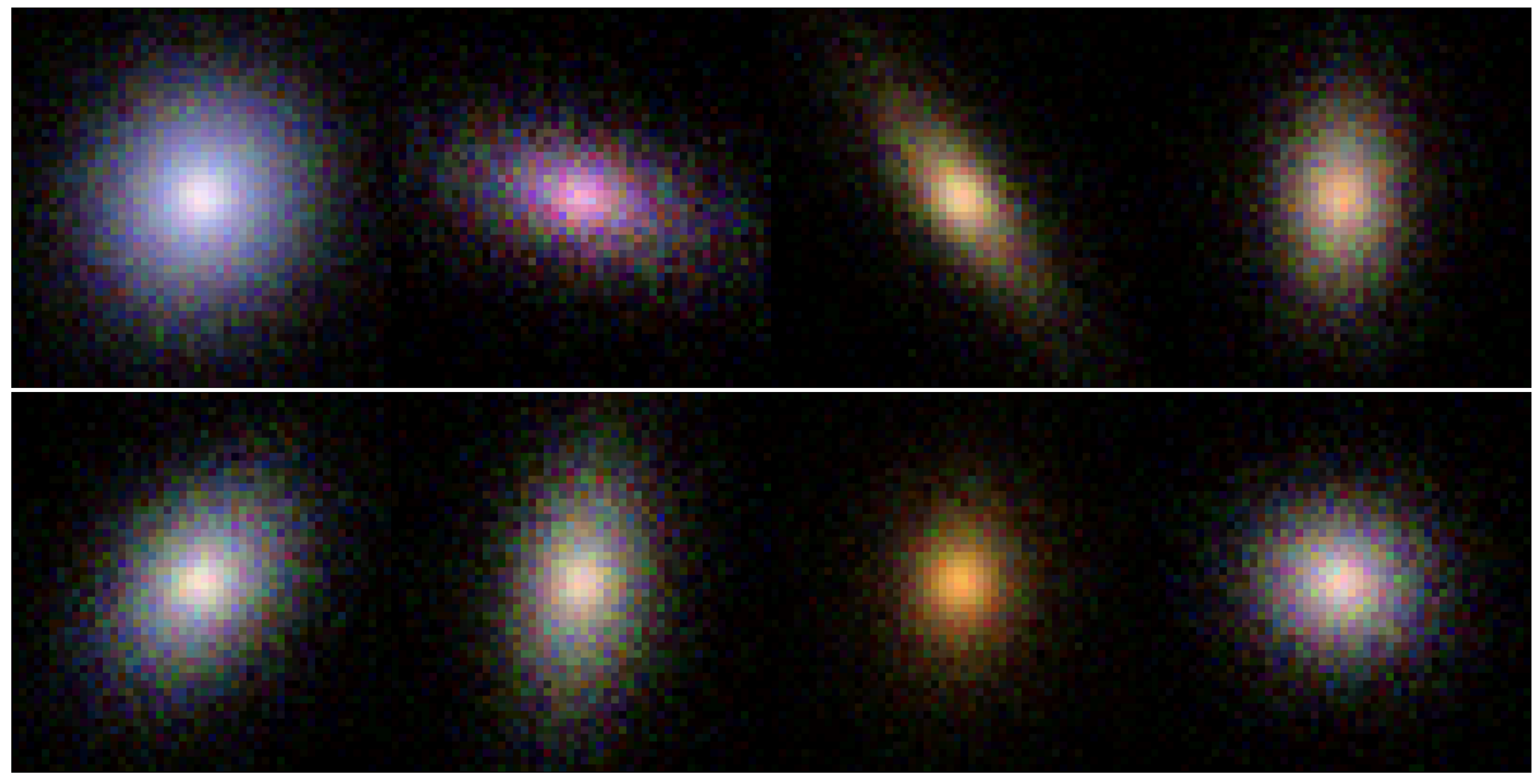}
    \caption{Randomly selected galaxies from our synthetic dataset.}
    \label{fig:sim_gals}
\end{figure}

The pixel resolution was set to 0.2 arcsec and the brightness distributions have been convolved with a Gaussian PSF having a FWHM of 0.7 arcsec in all filters. Finally, we added a simple Poisson noise on the 50x50x4 images to get more realistic measurements. In Figure \ref{fig:sim_gals} we plotted some of the synthetic galaxies for illustration purposes.

\section{Methods}\label{methods}

\subsection{Pre-processing of the data}

\begin{figure}
	\includegraphics[width=\columnwidth]{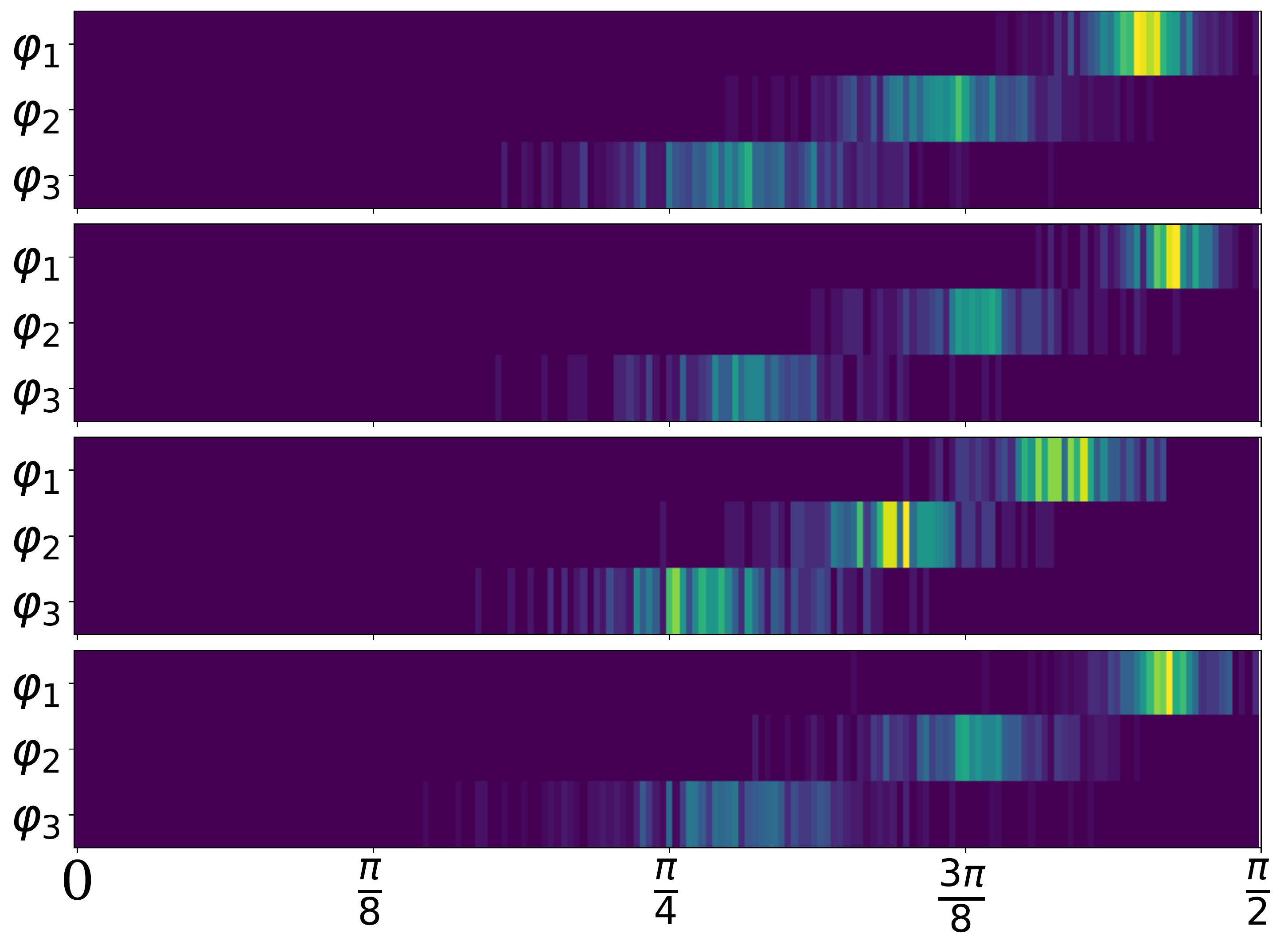}
    \caption{Color distribution in the 3 angle coordinates of four randomly selected galaxies. Colors indicates the number density in the bins.}
    \label{fig:color_dist}
\end{figure}

\begin{figure}
	\includegraphics[width=\columnwidth]{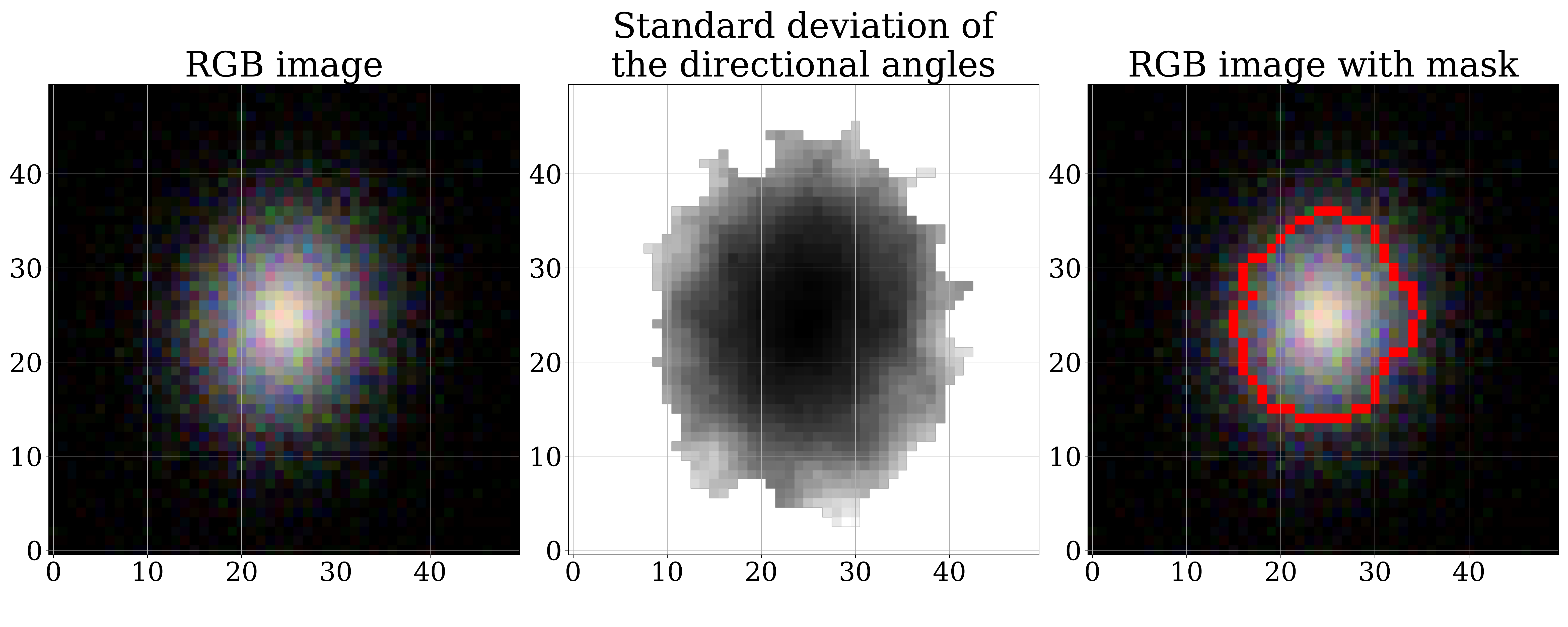}
    \caption{Mask determination using the color noise map.}
    \label{fig:mask_det}
\end{figure}

In order to provide the most relevant data for the SED predicting neural network (\texttt{sedNN}) we extracted the \textit{color} distribution of the galaxy inside a predefined mask. In this context \textit{color} means the directional angle of the pixel vectors in the 4 dimensional magnitude space. Since the pixel values of the sources should be positive, their angle coordinates will be in the range of $\varphi_1, \varphi_2, \varphi_3 \in [0,\pi/2]$. We divided this range into 180 bins, hence we ended up with 3x180 matrices (see Figure \ref{fig:color_dist}), which were later flattened for the neural network. It is worth mentioning that these matrices are \textit{independent from the exact morphology}, they only show the statistics of the pixel colors.\\
For the mask determination we calculated first a color noise map, where the standard deviation of colors were calculated inside a 5x5 sliding window (see Figure \ref{fig:mask_det}). We made a cut in the color noise map at 0.2 radian above which mostly the noisy edge of the galaxy can be found and which would debase the quality of the color histograms. 

\subsection{Network architecture of \texttt{sedNN}}

First we started with a network architecture of three fully connected hidden and dropout layers, one input and one output layer. All layers use elu activation function to avoid vanishing gradients. For training we used ADAM optimizer with a learning rate of 0.001, mean squared error loss function and a batch size of 128. To find the best performing structure we applied a grid search over two important hyperparameters, namely the dropout rate and the number of neurons in each hidden layer. We found that using a grid of \textit{number of neurons}x\textit{dropout rate}=[80,100,120,140]x[0.1,0.2,0.3,0.4,0.5] the best performing network has 120 neurons and a dropout rate of 0.2. We summarized these properties in Table \ref{tab:network_params}.\\
In the output layer we restricted the network to give such SEDs, which sum up exactly to one. This criterion was achieved by predicting only the first 3 components of the SED vectors (hence, the output dimension was 2x3=6) and calculating the last one by subtracting them from one. The network was then trained on these 8-element vectors containing the SEDs of the bulge and disk components.

\begin{table}
	\centering
	\caption{Hyperparameters of \texttt{sedNN} defining its architecture and the training process.}
	\label{tab:network_params}
	\begin{tabular}{cc} % four columns, alignment for each
		\hline
		Parameter & Value\\
		\hline
		Number of hidden layers & 3 \\
		Number of neurons per layer & 120 \\
		Dropout rate & 0.2 \\
		Activation function & elu \\
		Optimizer & Adam \\
		Loss function & Mean Squared Error \\
		Batch size & 128 \\
		Initial learning rate & 0.001 \\
		\hline
	\end{tabular}
\end{table}

\section{Results}\label{results}

\subsection{Predicting the SEDs}

\begin{figure}
	\includegraphics[width=\columnwidth]{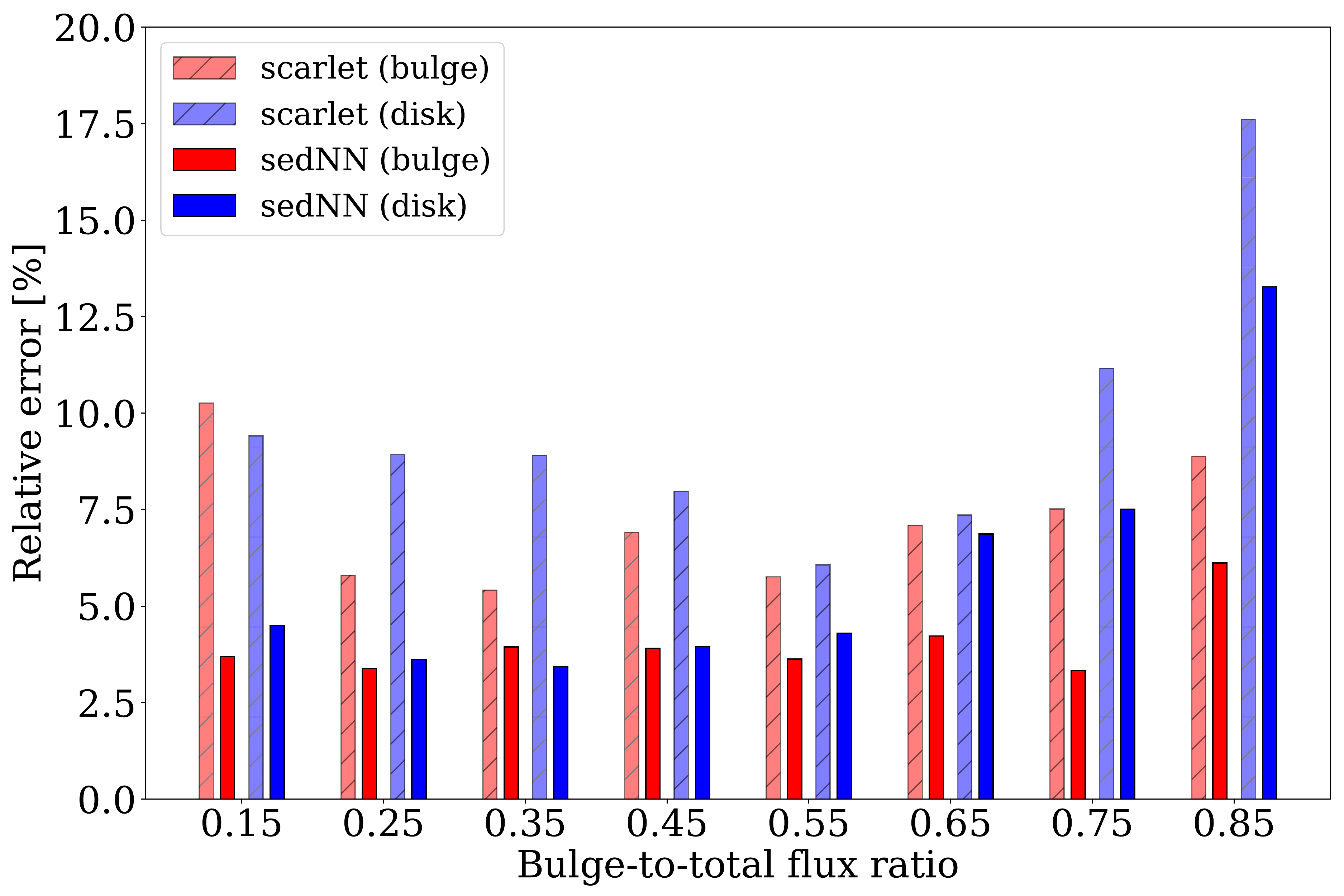}
    \caption{Relative error in the SED determination. Results are compared with \texttt{SCARLET}.}
    \label{fig:sed_res}
\end{figure}

 We evaluated the \texttt{sedNN} model performance on the test set grouped by the bulge-to-total flux ratios in 0.1 wide bins. For comparison we applied \texttt{SCARLET} on this subset as well, where we found that at all bulge-to-total flux ratios \texttt{sedNN} provided more accurate results (see Figure \ref{fig:sed_res}). We used the \texttt{MultiExtendedSource} class of \texttt{SCARLET}, where we gave the center of the object and the original PSF, and we set the number of components to $K=2$. In the simultaneous fitting process of the SEDs and morphology we used a maximum of 600 iteration and \texttt{e\_rel}$=10^{-6}$.\\
 We calculated the relative root mean squared error according to $RelRMS = RMS(SED_{original}, SED_{predicted})/SED_{original}$. We can see that on average the prediction efficiency of \texttt{sedNN} is between 3.3-6.1\% and 3.4-13.2\% for the bulge and disk, respectively. Contrarily, in case of \texttt{SCARLET} these errors are between 5.4-10.2\% and 6.1-17.6\%, which is about two times worse than \texttt{sedNN}. It is remarkable that in case of only 0.1 flux contribution of the bulge \texttt{sedNN} is able to predict its SED with 3.7\% as against to \texttt{SCARLET} 10.3\% efficiency. This indicates that one can more effectively predict the SEDs using the color distribution of the pixels than solving the whole source separation problem. The SED estimation performance of the disk is however worse in case of its low flux contribution (at $BT \approx 0.9$), which is mainly due to the low signal-to-noise ration near the edge of the galaxy. Nevertheless, even in this case \text{sedNN} prediction is 5\% better than using \texttt{SCARLET}.
 
 \subsection{Deblending the galaxies}
 
 Since we have compared our results to \texttt{SCARLET}, in this study we also tested the deblending performance of \texttt{SCARLET} on the morphology determination of the bulge and disk. In the following we present the correlation between the derived and original components according to the same equation used in \citet{scarlet} (see Eq. \ref{eq:corr}) as well as the relative flux error of the components.
 
 \begin{equation}
     \zeta_\nu = \frac{\nu_{true} \nu_{observed}}{\sqrt{\nu_{true}\nu_{true}}\sqrt{\nu_{observed}\nu_{observed}}}
     \label{eq:corr}
 \end{equation}
 
 We calculated these measures for the original deblending -- where the SED and the morphology are simultaneously fitted -- and we also repeated the source separation process while fixing the SEDs of the components to the more accurate values predicted by \texttt{sedNN}. In \texttt{SCARLET} by default the total flux of the components in each band is fitted through the SEDs and the morphology matrices are restricted with a normalization constraint. Since we only provide the normalized SED we had to make some modifications to fit the total fluxes through the morphology. In the following we have compared the PSF convolved brightness distributions to avoid potential systematic effects of the deconvolved images.
 
\begin{figure}
	\includegraphics[width=\columnwidth]{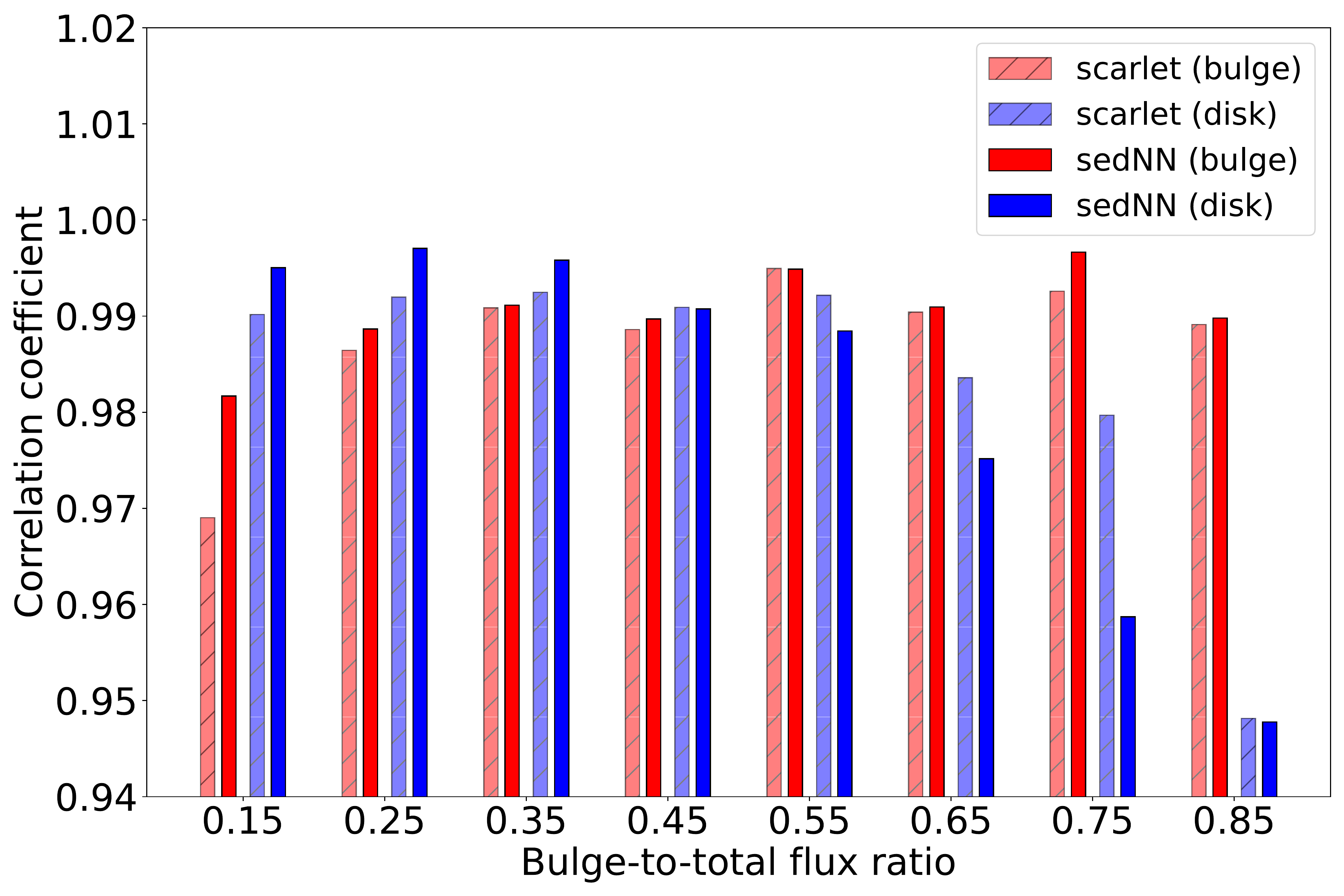}
    \caption{Correlation between the derived and the true morphologies of the galaxy components.}
    \label{fig:corr}
\end{figure}

\begin{figure}
	\includegraphics[width=\columnwidth]{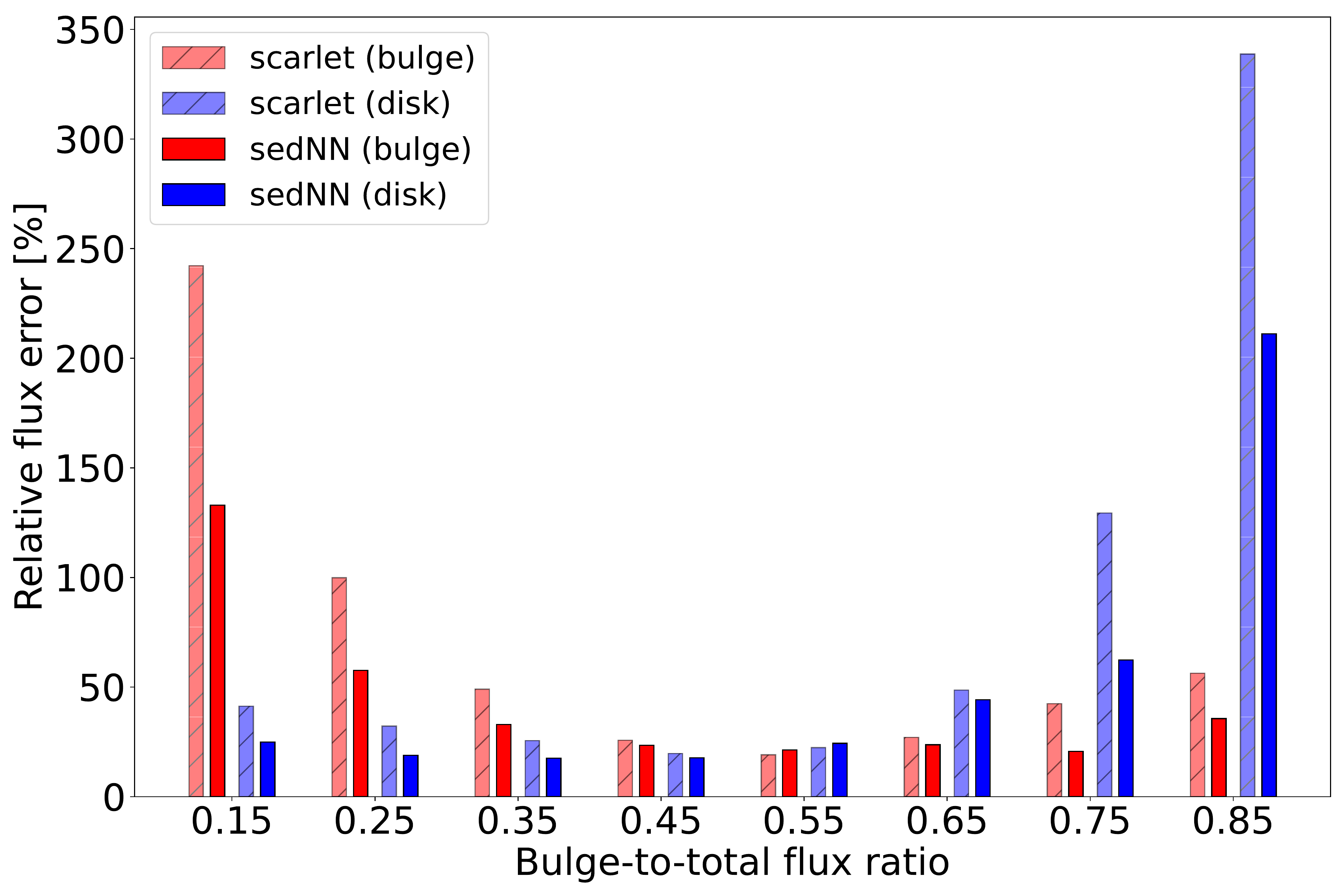}
    \caption{Relative error in the flux determination of the components.}
    \label{fig:fluxerr}
\end{figure}

 In Figure \ref{fig:corr} we plotted the correlation coefficients at the different bulge-to-total flux ratios. We can observe that the better SEDs improve the morphology determination mostly below $BT < 0.5$ and near to the two edges of the bulge-to-flux ratios. In Figure \ref{fig:fluxerr} we plotted the relative error in the flux determination of the components. The improvement is now more significant where the relative error is about 30\% smaller if we provide the more accurate SEDs and fix them during the source separation process. These results clearly shows us that \texttt{SCARLET} performs well in the simultaneous fitting if the two components are equally bright. However, in situations where the bulge is less dominant (e.g. in irregular galaxies) or the disk is very faint (e.g. old spiral galaxies with a large bulge) the improvement of the SED prediction results in a significant improvement in the flux estimation as well.
 
\section{Conclusions}\label{conclusions}

In our work we developed a \textit{morphology-independent} method to estimate the normalized broadband spectral energy distributions of the two main stellar populations in simulated disk galaxies, where we exploit the relation between the observable color distribution of the pixels and the real SEDs. We built a neural network (\texttt{sedNN}) which has been successfully trained on realistic simulated galaxies of CosmoDC2 to predict the real SEDs of the blue and red stellar population even if the flux contribution of them is very different. These accurate estimations were further given to \texttt{SCARLET}. We have shown that in the case of the used simulated galaxies our two-step approach has in most cases better deblending performance than simultaneously fitting the SEDs and the morphology with \texttt{Scarlet}, although the reproduction error of the components is still high. This shows us how difficult is the model-independent bulge-disk separation of spiral galaxies even in case of a synthetic galaxy catalog.\\
The developed model could be applied for real observations of spiral galaxies where we could use Integral Field Spectroscopy measurements to create the ground truth values for the broadband spectral energy distributions. 

\section*{Acknowledgements}

This work was supported by the  Ministry of Innovation and Technology NRDI Office grants OTKA NN 129148 and the MILAB Artificial Intelligence National Laboratory Program.

%%%%%%%%%%%%%%%%%%%%%%%%%%%%%%%%%%%%%%%%%%%%%%%%%%
\section*{Data Availability}

We used data from the public archives of CosmoDC2 galaxy catalog\footnote{ref: https://portal.nersc.gov/project/lsst/cosmoDC2/}. The further processed data as well as the model architecture of \texttt{sedNN} are also available at https://github.com/skunsagimate/sedNN.

%%%%%%%%%%%%%%%%%%%% REFERENCES %%%%%%%%%%%%%%%%%%

% The best way to enter references is to use BibTeX:

\bibliographystyle{mnras}
\bibliography{references} % if your bibtex file is called example.bib

% Alternatively you could enter them by hand, like this:
% This method is tedious and prone to error if you have lots of references
%\begin{thebibliography}{99}
%\bibitem[\protect\citeauthoryear{Author}{2012}]{Author2012}
%Author A.~N., 2013, Journal of Improbable Astronomy, 1, 1
%\bibitem[\protect\citeauthoryear{Others}{2013}]{Others2013}
%Others S., 2012, Journal of Interesting Stuff, 17, 198
%\end{thebibliography}

%%%%%%%%%%%%%%%%%%%%%%%%%%%%%%%%%%%%%%%%%%%%%%%%%%

%%%%%%%%%%%%%%%%% APPENDICES %%%%%%%%%%%%%%%%%%%%%

%%%%%%%%%%%%%%%%%%%%%%%%%%%%%%%%%%%%%%%%%%%%%%%%%%

% Don't change these lines
\bsp	% typesetting comment
\label{lastpage}
\end{document}